# Biomimetic Micropillar Wick for Enhanced Thin-Film Evaporation

*Anand S and Chander Shekhar Sharma\**

Thermofluidics Research Lab, Department of Mechanical Engineering,
Indian Institute of Technology Ropar, Rupnagar, Punjab 140 001, India
<chander.sharma@iitrpr.ac.in>, Ph: +91-1881-232358

## ABSTRACT

Sustainable liquid cooling solutions are recognized as the future of thermal management in the chip industry. Among them, the phase change heat transfer devices such as heat pipes and vapour chambers have shown tremendous potential. These devices rely on the physics of capillary-driven thin-film evaporation which is inherently coupled with the design and optimization of the evaporator wicks used in these devices. Here, we introduce a biomimetic evaporator wick design inspired by the peristome of the Nepenthes alata that can achieve significantly enhanced evaporative cooling. It consists of array of micropillars with multiple wedges along the sidewall of each micropillar. The efficacy of the wedged micropillar is evaluated based on a validated numerical model on the metrics of dryout heat flux and effective heat transfer coefficient. The wedge angle is chosen such that wedged micropillars cause liquid filaments to rise along the micropillar vertical walls. This results in a significant increase in thin-film area for evaporation. Additionally, the large mean curvature of the liquid meniscus produces strong capillary pumping pressure and simultaneously, the wedges increase the overall permeability of the wick. Consequently, our model predicts that the wedged micropillar wick can attain ~234% enhancement of dryout heat flux compared to a conventional cylindrical micropillar wick of similar geometrical dimensions. Moreover, the wedged micropillars can also attain higher effective heat transfer coefficient under dryout conditions thus outperforming the cylindrical micropillar in terms of heat transfer efficiency. Our study provide insight into the design and capability of the biomimetic wedged micropillars as an efficient evaporator wick for various thin-film evaporation applications.

# INTRODUCTION

The upsurge in the component density of microelectronics to meet the rising demand for high-performance processors has led to a conspicuous thermal management challenge for thermal engineers. The integration of these microprocessor chips into small form factor mobile devices has increased the operating temperature as well as temperature non-uniformity of the device due to the accumulation of parasitic heat[1]. The traditional heat dissipation technologies based on forced air and liquid cooling meet the cooling demand at the expense of high pumping power, noise, and bulkier assembly[2]. The advent of two-phase heat transfer devices such as heat pipes and vapour chambers have enabled sustainable cooling of electronics by dissipating excess heat through the mechanisms of thin-film evaporation and capillary pumping[3]. Here the design and optimization of the porous wick structures are recognized as vital for enhancing the thermal performance of these devices by maximizing the capillary pumping pressure while minimizing the viscous pressure penalty associated with flow of liquid through the wick[2,4–7].

Among the various types of wick structures, well-defined micro/nano structures have been extensively studied due to their precise control of the design parameters during fabrication as well as their ease to understand the physical mechanism of heat and mass transfer at the fundamental level[8]. Such wicks consist of an array of micropillars, and their performance has been investigated both experimentally and numerically for thin-film evaporation applications. Wicks with various micropillar geometries such as pyramidal[9], conical[9,10], pies and clusters[11], rectangular ribs[12,13], mushroom-shaped[14], catenoidal[15], and gradient cuboid[16] have been explored. Additionally wicks with non-uniform micropillar distribution[17] and segmented arrangement of micropillars[7] have also been reported. The extension of the thin-film region of the liquid meniscus have been highlighted to be the fundamental cause for the higher performance of the above geometries with regard to the cylindrical micropillars. In terms of modeling, multiple numerical models have been developed, both with and without the consideration of variation in meniscus shape to optimize the geometrical dimensions of microstructures and to predict the dryout heat flux and evaporator superheat[5,6,8,18–20].

The present work proposes a biomimetic microstructure for enhanced thin-film evaporation which is engineered based on a specific feature of Nepenthes alata (pitcher plant). The peristome surface of this plant has two-tier hierarchical structure consisting of parallel microgrooves with microcavities that have wedge corners. These wedge corners enable passive spreading of the liquid over the peristome surface which in effect is the capillary rise along the corner of a wedge formed from two plates as shown in the inset of Fig. 1a. Based on this, we propose a microstructure for evaporator wicks consisting of micropillars with wedges along the vertical wall of each pillar. We refer to these microfeatures as wedged micropillars.

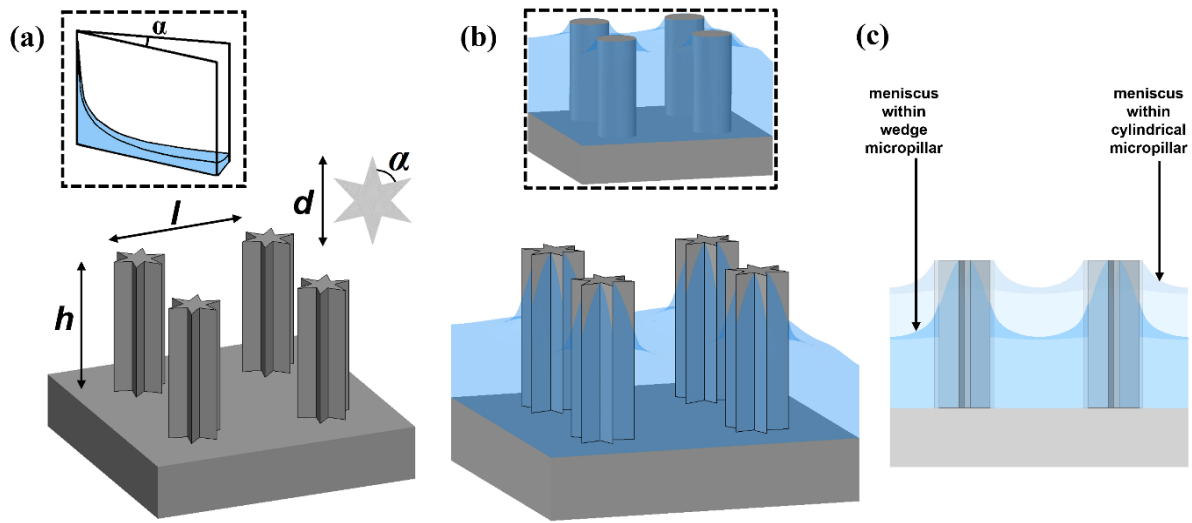

**Figure 1.** Schematic illustration of **(a)** biomimetic wedged micropillars showing geometrical parameters such as nominal diameter *d*, pitch *l*, height *h*, and wedge corner angle *α*. The inset image shows the capillary rise along the corner of a wedge formed from two plates which depicts the similar feature of the microcavity on the peristome upon which the biomimetic wedged micropillars are designed; **(b)** the liquid-vapor meniscus morphology within the biomimetic wedged micropillar array and the inset shows the same for conventional cylindrical micropillar; **(c)** side view highlighting the variation of meniscus curvature for both wedged and cylindrical micropillars of same geometrical parameters at $\theta=15°$ (refer Supporting Information, Section 1 for variation of meniscus curvature along the diagonal of the unit cell).

Figure 1a schematically illustrates the proposed morphology of micropillars, with each wedge having corner angle *α*, and the corresponding morphology of the liquid meniscus is shown in Fig. 1b. The cylindrical micropillar geometry of same nominal diameter, height and pitch and its meniscus morphology is shown in inset of Fig. 1b. It is well established that a wedge can cause a wetting liquid to rise along it when the wedge angle $\alpha$ is selected such that $\alpha + 2\theta \leq \pi$ where $\theta$ is the Young's contact angle of the liquid with the solid surface of the wedge[21]. We select the wedge angle accordingly such that a thin filament of liquid is maintained through capillarity along the wedge corner, thus extending the thin-film region for an enhanced rate of evaporation. As evident from Fig. 1c, the wedged micropillar causes a significant change in the meniscus morphology compared to the conventional cylindrical micropillar geometry for the same contact angle. In essence, while the wedges on the peristome of Nepenthes alata allow spreading of liquid along the surface, the wedges along the height of micropillars in the proposed microstructure enable liquid rise normal to the overall surface. We use a validated numerical approach to analyse the fluid transport and heat transfer performance of this biomimetic wedged micropillar structure in terms of dryout heat flux and effective heat transfer coefficient, which are the two key metrics for analysing the thermal performance of evaporator wicks. The dryout heat flux represents the maximum heat flux that is dissipated from the evaporator beyond which the reservoir cannot replenish the evaporator wick with the coolant. It depends on the capillary pumping pressure and viscous pressure drop for flow of the liquid coolant through the wick. The effective heat transfer coefficient characterizes the efficiency of heat dissipation through evaporation, and it depends on the extent of the thin-film region of the liquid meniscus. A significant part of total evaporation is concentrated within the liquid meniscus of few microns thickness and hence, enhancing the thin-film region leads to minimal interfacial and thin-film conduction resistances thus resulting in higher effective heat transfer coefficient[15,22]. Our results show that wedged micropillars with optimal morphology have the potential to achieve significantly higher passive thin-film evaporative cooling performance compared to the state-of-the-art wick designs.

## RESULT AND DISCUSSION

**Micropillar geometries**

The performance of the biomimetic wedged micropillar geometry is discussed in detail with respect to permeability of the wick, the dryout heat flux and the effective heat transfer coefficient for a micropillar array of length *L = 5 mm*, a nominal micropillar diameter *d = 20 μm,* and height *h = 50 μm*. The interstitial pitch *l = 30 μm* and *50 μm* (i.e., pitch to diameter (*l/d*) ratios of 1.5 and 2.5 respectively) are studied to understand the effect of the density of micropillars. We also analyse the effect of number of wedges along the circumference of the micropillar on thin-film evaporation. Overall, the thin-film evaporation performance of two wedged micropillar geometries, one with four wedges and the other with six wedges around the circumference, is compared against that of conventional cylindrical micropillars.

**Hydrodynamic performance**

Figure 2a compares the midplane averaged velocity as a function of the pressure gradient across a unit cell for the wedged and cylindrical micropillar geometries, for an exemplar contact angle of 50°. For more details on the estimation of midplane averaged velocity from the velocity magnitude contour obtained from the fluid flow model refer Supporting Information, Section 2. It is observed that for any given pressure gradient, all the three micropillar geometries achieve a higher midplane averaged velocity for sparse micropillar array (*l/d = 2.5)* compared to dense array (*l/d = 1.5*). This is expected as a denser array of pillars translates to a higher hydrodynamic resistance to the flow of the liquid through the wick. More interestingly, we find that the two wedged micropillar geometries achieve higher midplane velocities compared to cylindrical micropillar for both the sparse and dense arrays, thus indicating that the wedged micropillars offer a lower hydrodynamic resistance for wicking of liquid.

The above trends can be understood by considering the permeability of the flow for the given unit cells. Here permeability signifies the ease of flow within the micropillar array, estimated based on the Darcy's equation ($K = \frac{\mu U}{dp/dx}$). Figure 2b shows permeability of all three micropillar geometries as function of contact angle and *l/d* ratio. Three distinct aspects can be discerned from Fig. 2b. (a) First, for any micropillar geometry, a higher *l/d* ratio leads to higher permeability due to increased area available for liquid flow. This is also evident from the comparison of unit cell schematics for that micropillar geometry for high and low *l/d* ratio as shown in Fig. 2c-e. (b) Second, the wedged micropillars achieve higher permeability compared to cylindrical micropillars at both *l/d* ratios. This can be attributed to increased area of flow for wedged micropillars resulting from a lower solid fraction (defined as $f_s = \frac{volume\ of\ solid}{total\ volume\ of\ unit\ cell}$) compared to the cylindrical micropillar. For instance, $f_s$ for six-wedge micropillar is about 62% lower than that for cylindrical micropillars ($f_s$ values for all micropillar geometries are listed in Supporting Information, Section 3). The increased area available for flow in case of wedged micropillars is also evident on comparison of unit cell schematics across the three micropillar geometries for any *l/d* ratio from Fig. 2c-e. (c) Third, while the permeability for cylindrical micropillars shows no dependence on contact angle, the permeability for wedged micropillars reduces at lower contact angles for the case of higher *l/d* ratio. This is a direct consequence of the significant increase in curvature of the liquid meniscus for wedged micropillars at lower contact angles thereby reducing the cross-sectional area available for the flow. However, the wedged micropillars offer an overall lower hydrodynamic resistance compared to cylindrical micropillars for liquid wicking due higher permeability at most contact angles for low as well as high *l/d* ratio.

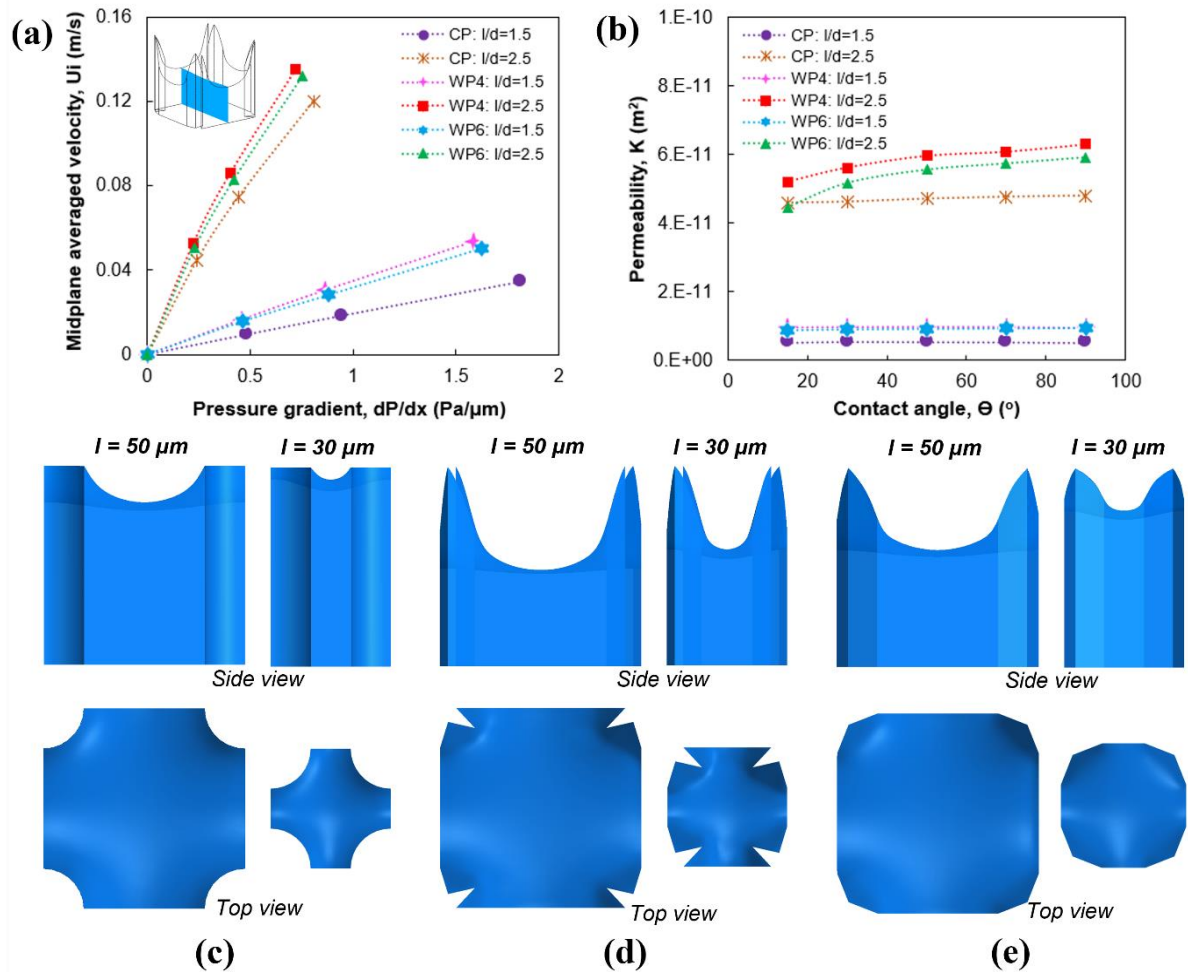

**Figure 2.** **(a)** Midplane (blue plane in the inset) averaged velocity as a function of pressure gradient for various types of micropillar geometries with $\theta=50^o$; **(b)** Variation of permeability of flow in relation to contact angle for various types of micropillar geometries. Here the legends CP, WP4, and WP6 refers to cylindrical, four-wedge, and six-wedge micropillars respectively. The geometric dimension of the micropillars is $d=20\mu m$, and $h=50\mu m$. Side and top view illustrations of sparse and dense array unit cell at $\theta=15^o$ of **(c)** cylindrical micropillar; **(d)** six-wedge micropillar; **(e)** four-wedge micropillar. The meniscus of six-wedge micropillar has more curvature followed by four-wedge and cylindrical micropillars. Both four-wedge and six-wedge micropillars have low solid fraction because their profile is obtained by the removal of material from the cylindrical micropillar.

**Thermal performance**

Figure 3a compares the predicted dryout heat flux, $q''_{dryout}$, for the biomimetic wedged micropillar arrays with cylindrical micropillar array, for the same diameter, height, and wicking length. The wedged micropillars show significantly higher $q''_{dryout}$ values compared to cylindrical micropillars at both low and high $l/d$ ratios. Additionally, the dryout heat flux increases with an increase in $l/d$ ratio for all micropillar geometries, which can be attributed to the increased permeability for the flow through the micropillar array. Overall, the six-wedge micropillar array with an $l/d$ ratio of *2.5* is predicted to achieve the highest $q''_{dryout}$ of 688.2 W/cm$^2$ among all micropillar geometries, which is an enhancement of 234% over the cylindrical micropillar array with same $l/d$ ratio. This is the outcome of increased Laplace pressure gradient for wedged micropillars as evident from Fig. 3b, which results in a higher capillary pumping pressure for these geometries. This is due to the higher curvature of meniscus in case of wedged micropillars because of the liquid rise along the corners of wedges.

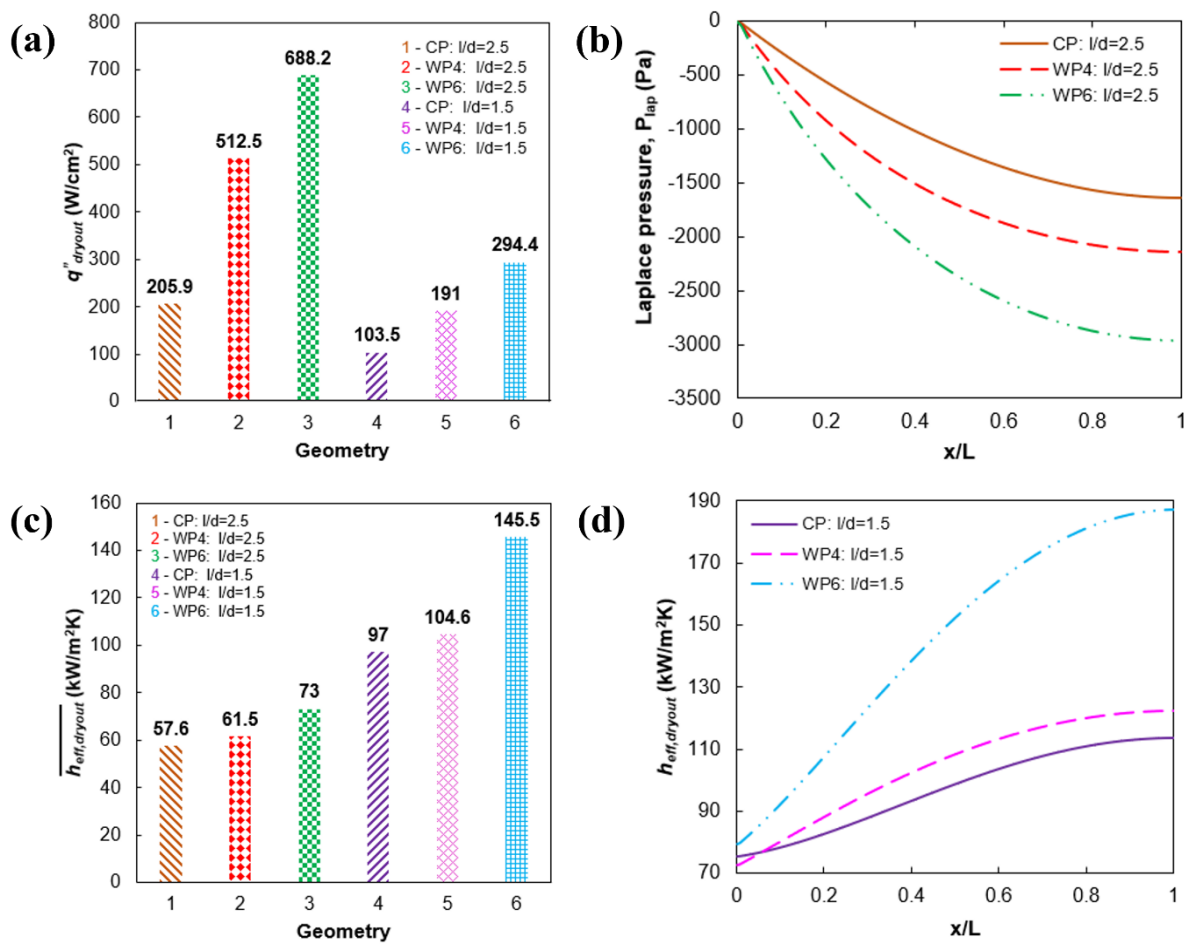

**Figure 3.** (a) Dryout heat flux for various types of micropillar geometries with *d = 20 μm, h = 50 μm, and L = 5 mm* at *l = 30 μm and 50 μm*; Number on top of each bar represents dryout

heat flux value for that geometry. **(b)** Laplace pressure variation along the flow direction for various types of micropillar geometries at *l/d = 2.5*. **(c)** Average effective heat transfer coefficient over the entire wicking length at the moment of dryout for different micropillar geometries (i.e. corresponding to the respective dryout heat fluxes of the three geometries). **(d)** Variation of effective heat transfer coefficient for dense array *l/d = 1.5* along the wicking direction at the moment of dryout for three micropillar geometries (corresponding plot for *l/d = 2.5* is shown in Fig. S3). Here the legends CP, WP4, and WP6 refers to cylindrical, four-wedge, and six-wedge micropillars respectively.

The second metric upon which the biomimetic wedged micropillars are evaluated is the effective heat transfer coefficient at the moment of dryout ($h_{eff,dryout}$). Figure 3c compares the average heat transfer coefficient over the entire wicking length at the moment of dryout, ($\overline{h_{eff,dryout}}$) for the three micropillar geometries. As seen earlier in Fig. 3a, the dryout heat flux is higher for sparse (*l/d = 2.5*) micropillar array for any micropillar geometry. In contrast, $\overline{h_{eff,dryout}}$ is higher for the dense (*l/d = 1.5*) array due to its lower overall thermal resistance. For instance, the six-wedge micropillar has $\overline{h_{eff,dryout}}$ values of *73 kW/m².K* and *145.5 kW/m².K* for sparse and dense arrays respectively. Still, overall, the six-wedge micropillars show the highest $\overline{h_{eff,dryout}}$ values followed by four–wedge micropillars. The cylindrical micropillars have the lowest $\overline{h_{eff,dryout}}$ value of *57.6 kW/m².K* and *96.96 kW/m².K* for sparse and dense arrays.

Figure 3d shows the variation of $h_{eff,dryout}$ along the wicking length for the respective dryout heat fluxes corresponding to dense arrays of the three micropillar geometries. It is evident that $h_{eff,dryout}$ for any micropillar geometry increases along wicking length. This is because the meniscus curvature increases due to evaporation. $h_{eff,dryout}$ variation is similar for sparse arrays. With $h_{eff,dryout}$ known, the evaporator temperature variation along the wicking length can be calculated by using equation (7) (refer Supporting Information section 4 for further details.) When the three micropillar geometries are compared for same heat flux load, the performance in terms of $h_{eff}$ is similar across the geometries (refer Supporting Information section 5 wherein $h_{eff}$ is compared across the three micropillar geometries at dryout heat flux for cylindrical micropillars). In essence, the wedged micropillar geometry dissipates heat with similar efficiency as cylindrical micropillars (due to comparable heat transfer coefficients) at lower heat loads. However, it can not only sustain significantly higher heat loads before dryout but can also dissipate such higher heat loads with higher efficiency (due to higher heat transfer coefficients under dryout condition) compared to cylindrical micropillars.

The overall higher heat transfer coefficient for wedged micropillars is the result of lower overall thermal resistance for this micropillar geometry compared to cylindrical micropillars. The overall thermal resistance is composed of (i) conduction resistances of micropillar, bulk liquid, and thin-film region and (ii) interfacial resistances across thin-film region and bulk meniscus as detailed in the Supporting Information, section 6. The conduction resistance through the bulk liquid is notably reduced for wedge micropillars due to the higher curvature of the liquid-vapour interface as shown in Fig. 2c-e. The interfacial resistance on the other hand depends on the extent of thin-film area available for evaporation. This is because a significant portion of evaporation is concentrated within this thin-film area of the liquid meniscus. We define thin-film area for any given meniscus as the region of the meniscus with thickness of the film as 5 $\mu m$ or smaller[7,12]. Figure 4a illustrates the thin-film region around an individual six-wedge micropillar based on this criterion of liquid film thickness. We quantified the area of this thin-film region for all micropillar geometries in terms of thin-film area fraction defined as the ratio of thin-film area to the projected area of the unit cell. Figure 4b shows the thin-film area fraction for all micropillar geometries corresponding to the dryout condition. The six-wedge micropillar achieves the highest thin-film area fraction followed by four-wedge and cylindrical micropillars. Also, dense array outperforms the sparse array in terms of the thin-film area fraction. This is consistent with the $\overline{h_{eff,dryout}}$ values shown in Fig. 3c. Thus the biomimetic wedged micropillar geometry can be designed to achieve significant enhancement in dryout heat flux and effective heat transfer coefficient.

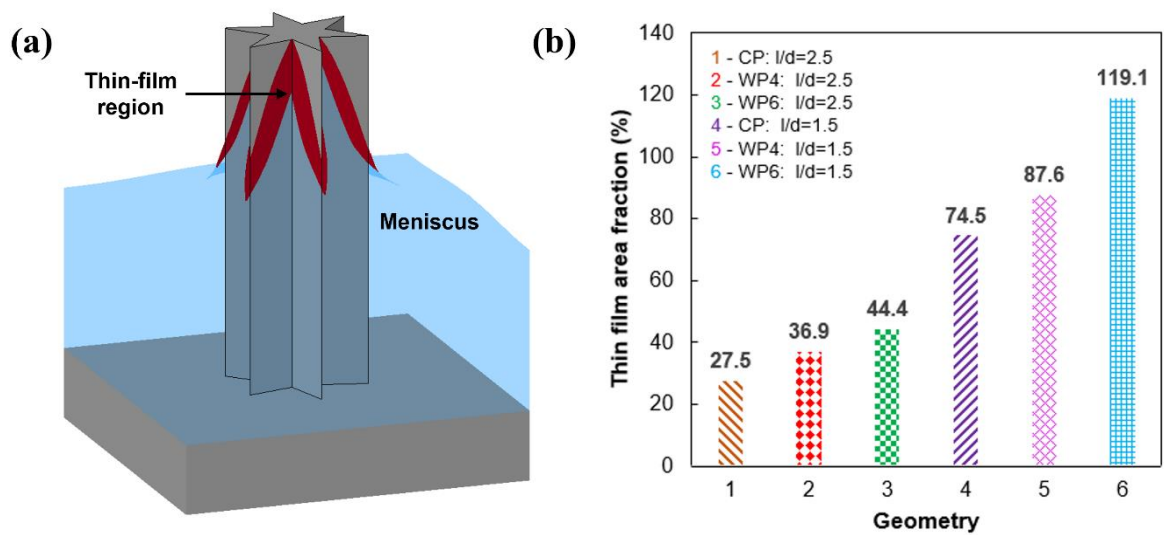

**Figure 4. (a)** Schematic illustration of the thin-film region of the liquid meniscus formed within a six-wedge micropillar; **(b)** Thin-film area fraction (defined as the ratio of thin-film area to the projected area of the unit cell) at the moment of dryout for different micropillar geometries. Here CP, WP4, and WP6 refers to cylindrical, four-wedge, and six-wedge micropillars respectively.

The proposed exemplar micropillar geometries can be fabricated using the standard microfabrication techniques involving lithography and deep reactive ion etching. During fabrication, the sharp corners in the wedged micropillars can get rounded off with a small but finite radius of curvature off in the order of one micron based on the resolution of the optical lithography process which can affect the capillary rise and needs to be accounted for during practical implementation. On the other hand, the capillarity of the fabricated wedged micropillars can be further enhanced by conditioning the wettability state of the surface towards superhydrophilicity (reduced $\theta$) through suitable chemical treatments[2].

## NUMERICAL FRAMEWORK

The numerical modelling framework is developed to predict dryout heat flux and effective heat transfer coefficient[8,23] and it is an integration of unit cell level models and a device level model[19]. The unit cell model is used to compute the effect of meniscus shape and geometrical parameters, such as micropillar pitch and morphology, on pressure drop and heat transfer. To this end, the equilibrium meniscus shape is estimated as a function of microstructure geometry through the principal of total energy minimization. Subsequently, the pressure drop and heat transfer characteristic corresponding to each meniscus shape is obtained through three-dimensional fluid flow and thermal CFD simulations. Next, a device-level model is built to simulate the performance of the evaporator wick. The wick is discretised into a set of finite volumes or unit-cells which are then used to account for variation in meniscus shape, coolant velocity and pressure along the wicking direction of coolant through mass, momentum, and energy conservation to forecast the dryout heat flux. The unit-cell is defined in Fig. 5a. Additionally, the variation in the effective heat transfer coefficient along the flow direction is derived from the device level model and is used to calculate the evaporator surface temperature distribution.

**Equilibrium meniscus interface**

The calculation of liquid/vapour interfacial profile and its variation along the coolant flow direction is critical for the prediction of dryout heat flux and evaporator temperature[8,19,24,25]. On this line, we calculate the three-dimensional (3D) meniscus shape for a given unit cell by the use of Surface Evolver[26,27] to develop the equilibrium static meniscus shape based on the total energy minimization approach. The total energy balance is expressed in terms of surface energy ($E_s$) as shown in Equation (1). The potential energy due to gravity is neglected as a result of a low Bond number (~ $10^{-4}$).

$$E_s = \iint_{A_{lv}} \gamma_{lv}\, dA - \iint_{A_{sl}} \gamma_{lv} \cos\theta\, dA \tag{1}$$

Here $\gamma_{lv}$ is the liquid-vapour surface tension, $A_{lv}$ and $A_{sl}$ are the liquid-vapour and solid-liquid surface areas respectively. Surface evolver employs the gradient descent algorithm to evolve the meniscus towards a minimum energy state, with the force on each vertex determined as the negative gradient of the surface energy during each iteration step. We set the convergence condition so that the change in surface area during the *10* iteration steps is less than *0.01%* [22] to achieve the equilibrium meniscus for a unit cell as depicted in Fig. 5a. The resultant meniscus surface is imported as point clouds into CAD software for the creation of the 3D liquid domain, as illustrated in Fig. 5b, which is then employed in fluid flow and thermal analysis. A wide range of meniscus shapes are computed for contact angles between *90°* and *15°*. The device level model requires information on average pressure and flow area in each unit cell. Hence, average Laplace pressure difference, $P_{lap} = P_v - P_l$, corresponding to each meniscus curvature is obtained from the force balance Equation (2). Here $P_v$ and $P_l$ are the vapor and liquid pressure respectively.

$$-P_{lap} = \frac{\gamma_{lv} \cos\theta\, P_w}{A_{pm}} \tag{2}$$

where $P_w$ and $A_{pm}$ are the wetted perimeter and projected meniscus area respectively. As the contact angle made by the meniscus along the periphery of the wedged micropillar was found to vary significantly along the contact line, we accounted for this variation in the above calculation of the Laplace pressure (refer Supporting Information, Section 7 for further details of Laplace pressure calculation). For a given geometry of micropillar, the cross-sectional area of the midplane of the unit cell $A_i$, is considered as the available flow area for use in the device model. It is represented as a function of the shape of the meniscus ($P_{lap}$) as,

$$A_i = f(P_{lap}) \tag{3}$$

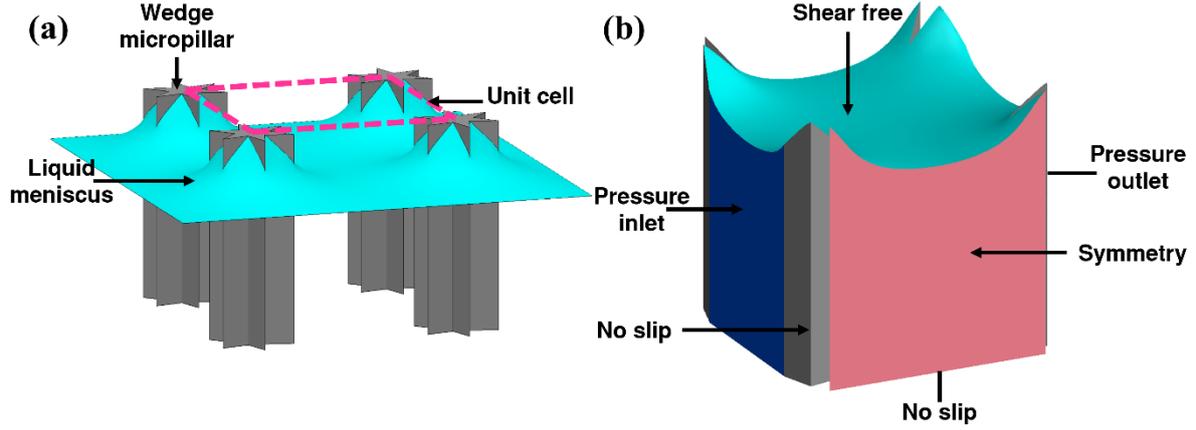

**Figure 5. (a)** Equilibrium shape of the meniscus generated from Surface Evolver highlighting a unit cell of wedged micropillar; **(b)** Boundary conditions applied in the unit cell fluid flow model include pressure inlet and outlet (dark blue), no slip at the unit cell base and pillar walls (grey), symmetric boundary at the two lateral faces (red), and shear-free surface on the liquid-vapour interfacial meniscus (cyan).

**Fluid flow model**

Three-dimensional fluid flow simulations are performed to compute the velocity distribution and pressure drop for flow of liquid coolant through the unit cell as a function of the shape of the meniscus. Conservation equations for mass (equation (4)) and momentum (equation (5)) are solved for the unit cell domain shown in Fig. 2b by using ANSYS FLUENT solver[28].

$$\nabla \cdot (\rho V) = 0 \quad (4)$$

$$\nabla \cdot (\rho V V) = -\nabla p + \nabla \cdot (\mu \nabla V) \quad (5)$$

Figure 2b displays boundary conditions such as pressure boundary conditions assigned at the inlet and outlet, no slip at the unit cell base and pillar walls, symmetric boundary at the two lateral faces, and shear-free surface on the liquid-vapour interfacial meniscus. The liquid's thermophysical properties are chosen corresponding to the saturation temperature of 100°C from NIST database[29]. The SIMPLEC technique is employed for pressure-velocity coupling for the creeping flow condition prevailing here in this case. The criteria for convergence for both continuity and momentum were set to $10^{-6}$ during the CFD simulation. Grid independent meshes are used throughout the unit cell simulations based on the grid independence check performed as detailed in the Supporting Information, Section 8. Using the unit cell solution data, average velocity ($U_i$) at the midplane of the unit cell is defined as a function of the shape of the meniscus ($P_{lap}$) and the pressure gradient ($\frac{dP}{dx}$) in the flow direction shown by equation (6),

$$U_i = f\left(P_{lap}, \frac{dP}{dx}\right) \tag{6}$$

For more details on the estimation of midplane averaged velocity refer Supporting Information, Section 2. Further, these data are used in the device level model for predicting the dryout heat flux.

**Heat transfer model**

For each unit cell, a steady-state thermal simulation is run to estimate the effective heat transfer coefficient ($h_{eff}$) due to evaporation based on the input heat flux ($q"$) and the evaporator wall superheat ($\Delta T_e = T_e - T_{sat}$) as given by equation (7) where $T_e$ is the temperature of evaporator surface at the solid-liquid interface and $T_{sat}$ is the saturation vapour temperature. The heat transfer model accounts for the thickness of the substrate (see Supporting Information section 9 for details) and heat conduction through the liquid domain dominates convection in this simulation (Peclet number ~ *10⁻¹*). Conservation equations (7-9) are solved for the conjugate heat transfer. It is considered that the heat flux from substrate is dissipated by evaporation from the liquid/vapour interfacial meniscus. The boundary conditions employed in this computational study include adiabatic condition at the pillar top surface, symmetry on all four lateral faces, convective heat transfer coefficient at the meniscus interface implemented through the Schrage's model for evaporation[30], and input heat flux at the substrate's bottom face as depicted in Figure S8a of Supporting Information section 9. The Schrage equation (9) gives the heat transfer coefficient ($h_{lv}$) for a case of evaporation dominant heat transfer, which is a fair supposition for many realistic thin-film evaporative cooling applications.

$$h_{eff} = \frac{q"}{\Delta T_e} \tag{7}$$

$$\nabla.[V(\rho C_p T)] = \nabla.(k\nabla T) \tag{8}$$

$$h_{lv} = \frac{2\sigma}{2-\sigma}\left(\frac{h_{fg}^2}{T_v v_{fg}}\right)\left(\frac{\bar{M}}{2\pi \bar{R} T_v}\right)^{1/2}\left[1 - \frac{P_v v_{fg}}{2h_{fg}}\right] \tag{9}$$

Here $\sigma$ denotes the evaporation accommodation coefficient which is taken to be *0.052*[8], $h_{fg}$, $v_{fg}$, $T_v$, $P_v$, $\bar{M}$, and $\bar{R}$ are the latent heat of vaporization, difference of vapour-liquid specific volume, vapour temperature, vapour pressure, molar mass, and universal gas constant respectively of the water evaluated at the saturation condition. Grid independent meshes are used throughout the unit cell simulations based on the grid independence check performed as detailed in the Supporting Information, Section 8. The effective heat transfer coefficient ($h_{eff}$) is then expressed in terms of the shape of the meniscus ($P_{lap}$) as

$$h_{eff} = f(P_{lap}) \qquad (10)$$

Along with the fluid flow data, equation (10) is fed into the device level model to obtain the distribution of the effective heat transfer coefficient along the flow [8,10].

**Device level model**

A one-dimensional finite volume discretization approach is adopted to develop the device level model for simulation of thin-film evaporation from the forest of wedged micropillar wick of length $L$ (Fig. 6a)[19]. Due to the periodic nature of the flow pattern in the $y$ direction, we essentially model only a single row of unit cells starting from $x = 0$ to $x = L$ of the wick, as highlighted in Fig. 6b. Here a single unit cell/finite volume encompasses the four quarters of micropillar and volume of liquid held within four pillars[19] (see Fig. 5a). We have assumed here that the system considered is at the saturation condition, hence throughout the domain vapour pressure $P_v = P_{sat}$. The microstructures withdraw liquid from reservoir through capillary wicking. As depicted in Fig. 6b, starting from reservoir (at $x = 0$), the meniscus curvature increases in the direction of flow (away from the reservoir) as a result of the evaporation. This translates to a reducing liquid pressure, $P_l(x)$ along the flow and is expressed as a relative pressure ($P_{lap}(x) = P_v - P_l(x)$) with respect to the vapour pressure. The resulting favourable pressure gradient along $x$ enables wicking of liquid coolant in that direction. As the heat flux is increased, the evaporation rate also increases throughout the wick which results in increase in meniscus curvature in all unit cells. Eventually, the meniscus curvature in the last unit cell reaches its maximum when the applied heat flux equals the dryout heat flux. This maximum in meniscus curvature corresponds to the local contact angle reducing to receding contact angle. Further evaporation causes the contact line to depin from the micropillar top edge and recede, leading to liquid dryout. Thus the contact angle in the last unit cell reducing to the receding contact angle becomes a criterion to detect the dryout heat flux[19]. In the model, water is considered as the working fluid and Silicon is considered as the solid material for which the receding contact angle is taken as *15°* throughout this modeling[31].

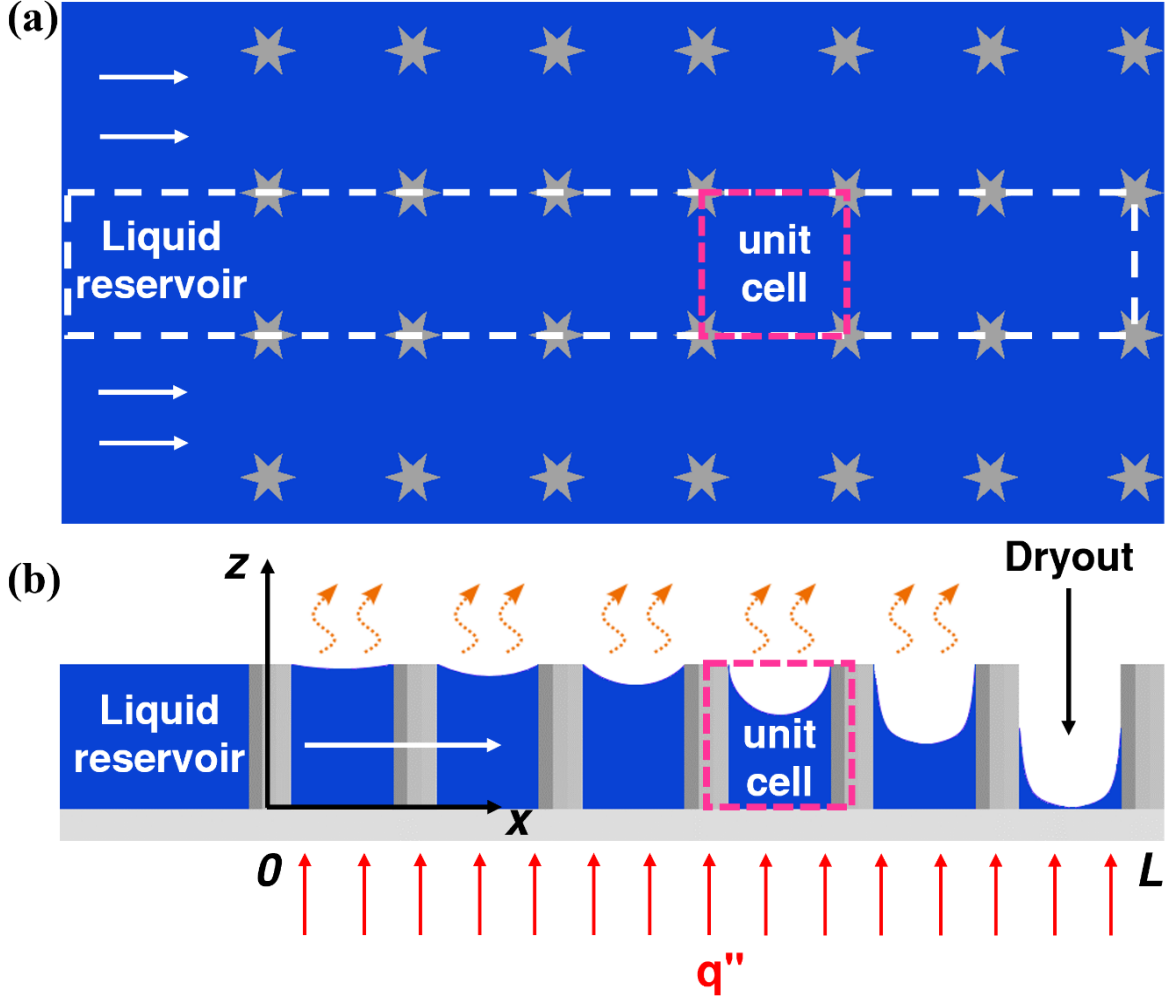

**Figure 6.** Schematic illustration of the forest of wedged micropillar wick highlighting the computational domain of the model where the liquid is driven from the liquid reservoir in the *x* direction (a) top view of array of length *L* and (b) side view of a single row of unit cells with streamwise variation of meniscus curvature and dryout in the last unit cell.

For a given geometry of the micropillar array and input heat flux condition ($q''$), the device level model couples each unit cell starting from the reservoir at *x = 0* to *x = L* based on the mass and enthalpy conservation at the midplane of the two adjacent unit cells. This results in equation (11) for enthalpy balance,

$$\rho(A_{i-1}U_{i-1} - A_i U_i)h_{fg} = q'' l^2 \qquad (11)$$

where $\rho$, $A_i$, and $U_i$ are the density, midplane area, and average velocity of the unit cell respectively. Equation (11) is next solved in an iterative method using an in-house MATLAB script after substituting equations (3) and (6), with the following boundary conditions: (a) at *x = 0*, $P_{lap_i} = 0$, i.e. zero mean curvature meniscus at the reservoir; (b) at *x = 0*, $\rho A_0 U_0 = \frac{q'' lL}{h_{fg}}$ is the assumed incoming mass of liquid which corresponds to the total mass evaporated from the entire row of the array. This results in the distribution of $P_{lap_i}$ and hence the $U_i$ along the flow direction. For predicting the dryout heat flux, the receding contact angle criteria is imposed as a limiting condition at the last unit cell ($x = L$) where the corresponding $P_{lap_i}$ approaches the maximum capillary pressure. From equation (10) the distribution of $h_{eff}$ as a function of wicking direction is applied as the convective boundary condition for evaporation in the device level model to obtain the evaporator base temperature ($T_e$) distribution (see Supporting Information, section 9 for details).

**Validation of the numerical framework**

The validation of the developed numerical framework is performed by modeling an evaporator wick with cylindrical micropillars with geometrical parameters of *d = 10 µm, l = 30 µm, h = 25 µm, t = 100 µm, and L = 5 mm*. The modeling results for the variation of Laplace pressure, dryout heat flux, effective heat transfer coefficient, and evaporator superheat demonstrate excellent agreement with the results of Zhu et al.[19] and Vaartstra et al.[8] as elaborated in the Supporting Information, Section 10.

# CONCLUSION

In this study, we have introduced a biomimetic wedged micropillar geometry, inspired from the peristome surface of Nepenthes alata, for cooling through capillary-fed thin-film evaporation. The performance of the wedged micropillar geometry has been evaluated by adopting a validated numerical modeling approach involving the integration of unit-cell based analysis into a one-dimensional device level model. The model involves generation of equilibrium meniscus shape using energy minimization approach and utilizing the resulting shape in unit cell-based fluid flow and thermal analysis. Subsequently, the results of unit cell-based analysis are integrated into a device level model by linking the unit cells for the prediction of dryout heat flux, average and local effective heat transfer coefficient, and evaporator wall superheat. Our modeling framework also accounts for the non-uniformity of contact angle along the contact line of the wedged micropillar while estimating the Laplace pressure. Using the model, the effect of density of micropillars as well as number of wedges along the micropillar periphery has been explored. In essence, two wedged micropillar geometries, with six and four wedges, are compared against conventional cylindrical micropillars. The significant improvement in the permeability of the flow for wedged micropillars, coupled with the higher capillary pressure, result in enhanced dryout heat flux for the wedged micropillars. Overall, a sparse array of six wedge micropillar attains the highest dryout heat flux which is ~234% higher than that a sparse array of the baseline cylindrical micropillars. Additionally, the higher thin-film area fraction and higher meniscus curvature leads to a higher effective heat transfer coefficient for wedge micropillars under dryout conditions compared to cylindrical micropillars.

## ASSOCIATED CONTENT

Meniscus curvature along the unit cell diagonal; Velocity contours and estimation of midplane averaged velocity; Solid fraction ($f_s$) of various micropillar geometries; Thermal performance of different micropillar geometries for the respective dryout heat flux; Thermal performance of different micropillar geometries for same heat flux; Thermal resistance network; Laplace pressure calculation for varying contact angles along the periphery of wedged micropillars; Grid independence check; Unit cell and device level heat transfer model; Validation of the numerical model.

## ACKNOWLEDGEMENTS

The authors gratefully acknowledge the funding provided for this work by Science and Engineering Research Board (SERB), Department of Science and Technology, Government of India under the Core Research Grant (CRG/2018/004539). We also acknowledge Prof. Kenneth A. Brakke of Susquehanna University for the insights provided on the Surface Evolver code and Siddhartha S. S. of Thermofluidics Research Lab, IIT Ropar for help with three-dimensional CAD profiles of the menisci.

# Biomimetic Micropillar Wick for Enhanced Thin-Film Evaporation

*Anand S and Chander Shekhar Sharma\**

Thermofluidics Research Lab, Department of Mechanical Engineering,
Indian Institute of Technology Ropar, Rupnagar, Punjab 140 001, India
chander.sharma@iitrpr.ac.in, Ph: +91-1881-232358

## TABLE OF CONTENTS

**SECTION 1: Meniscus curvature along the unit cell diagonal**

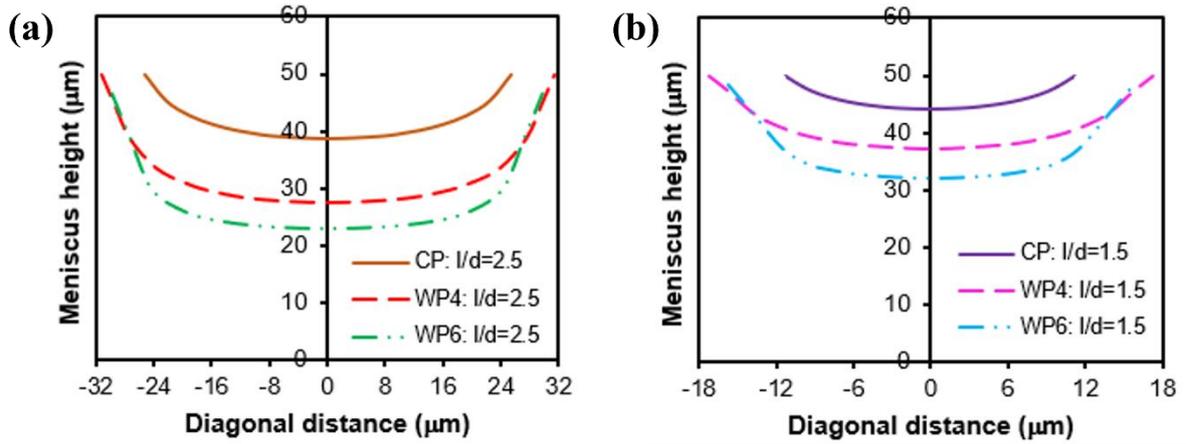

**Figure S1.** Variation of meniscus curvature along the diagonal of **(a)** sparse and **(b)** dense array unit cell of all three micropillar geometries.

The meniscus curvature variation along the diagonal of a unit cell for sparse and dense array of all three micropillar geometries at the moment of dryout is illustrated in Figure S1a and Figure S1b. The wedged micropillars demonstrates lower bulk liquid meniscus and higher corner rise than the cylindrical micropillars.

**SECTION 2: Velocity contours and estimation of midplane averaged velocity**

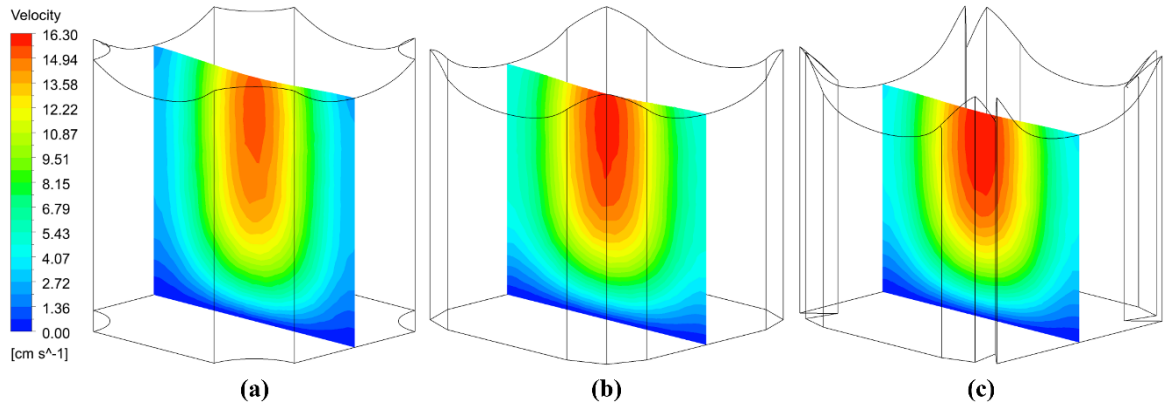

**Figure S2.** Velocity contours at the midplane of **(a)** cylindrical, **(b)** four-wedge, and **(c)** six-wedge micropillar geometries of *d=20 μm, l=50 μm, h=50 μm,* and *θ=50°* for a pressure drop of *30 Pa*

Based on the 3D unit cell fluid flow simulations, the velocity profile of the flow is obtained for the boundary conditions as mentioned in the main text. For the pressure driven flow, Figure S2 compares the unit cell velocity contours at the midplane of cylindrical, four-wedge, and six-wedge micropillar geometries of *d=20 μm, l=50 μm, h=50 μm,* and *θ=50°* for a pressure drop of *30 Pa*. The area averaged velocity ($U_i$) at the midplane of any given unit cell is computed as $U_i = \frac{1}{A_{MP}} \iint u(x,y,z)\, dydz|_{x=l/2}$, where $A_{MP}$ is the area of midplane, and $u$ is the *x* directional velocity.

## SECTION 3: Solid fraction ($f_s$) of various micropillar geometries

Solid fraction ($f_s$) is defined as the ratio of volume of the micropillar in a unit cell to the total volume of the unit cell.

**Table S1.** Geometrical parameters of various micropillar geometries

| Geometry | *d (μm)* | *l (μm)* | *h (μm)* | $f_s$ |
|---|---|---|---|---|
| Cylinder | 20 | 30 | 50 | 0.349 |
|  | 20 | 50 | 50 | 0.1256 |
| Four-wedge | 20 | 30 | 50 | 0.1257 |
|  | 20 | 50 | 50 | 0.0452 |
| Six-wedge | 20 | 30 | 50 | 0.134 |
|  | 20 | 50 | 50 | 0.048 |

## SECTION 4: Thermal performance of different micropillar geometries for the respective dryout heat flux

The variation of $h_{eff,dryout}$ along the wicking length was discussed in main manuscript for dense arrays and is reproduced in Figure S3a for reference. $h_{eff,dryout}$ variation is similar for sparse arrays as shown in Figure S3b. For instance, there is *~136%* and *~118%* change in $h_{eff,dryout}$ value from the reservoir (at *x/L=0*) to the centre (at *x/L=1*) of the array of six-wedge micropillars for dense and sparse arrays respectively. This leads to *~15 °C* and *~49 °C* difference in evaporator temperature between the reservoir and centre of the dense and sparse arrays respectively creating a temperature gradient along the wick as seen in Figure S3c and Figure S3d.

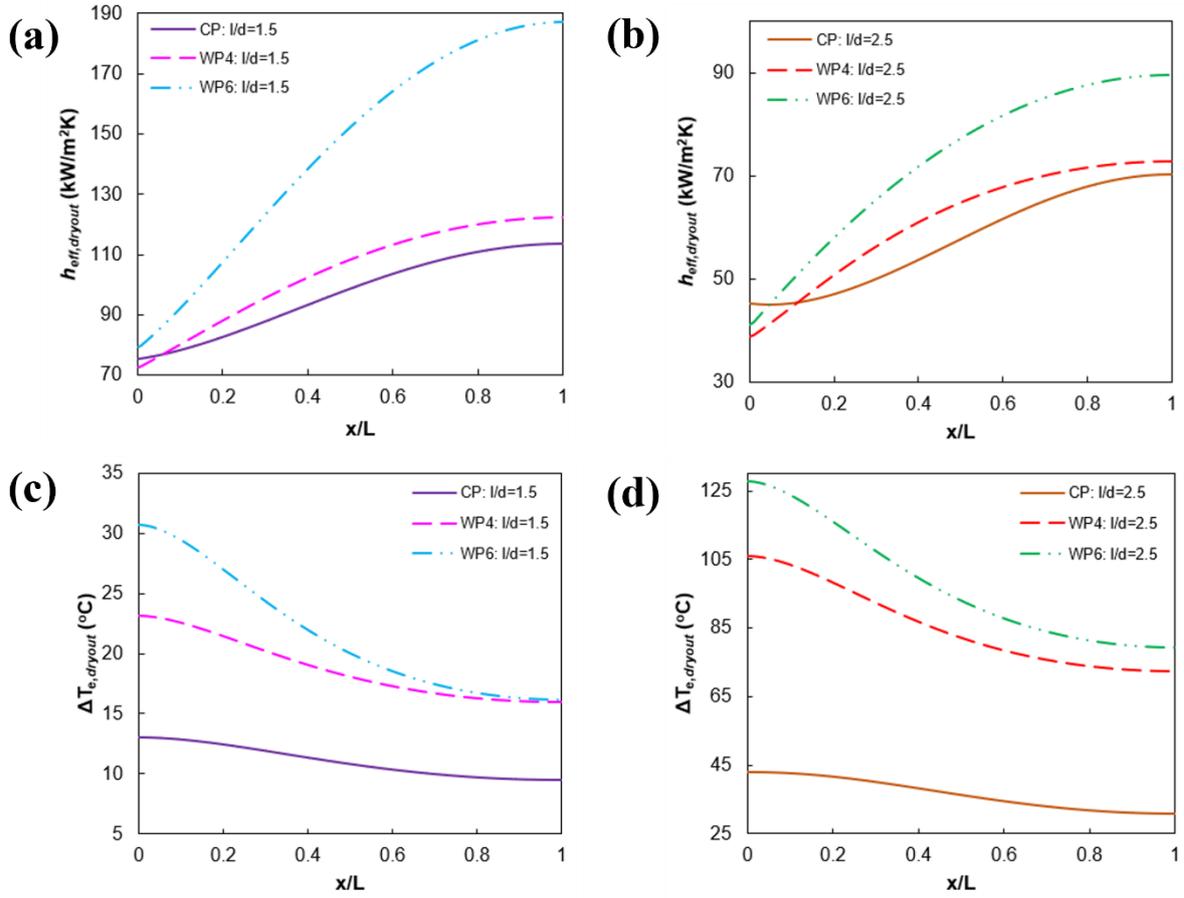

**Figure S3.** Variation of effective heat transfer coefficient for **(a)** dense and **(b)** sparse array along the wicking direction. Distribution of evaporator wall superheat for **(c)** dense and **(d)** sparse array along the wicking direction for all micropillar geometries at the respective dryout heat flux. Here CP, WP4, and WP6 refers to cylindrical, four-wedge, and six-wedge micropillars respectively.

**SECTION 5: Thermal performance of different micropillar geometries for same heat flux**

We analysed the thermal performance of the three micropillar geometries for same heat flux load which is equal to the dryout heat flux of dense cylindrical micropillar array (*103.5 W/cm$^2$*). The distribution of $h_{eff}$ and the evaporator wall superheat ($\Delta T_e$) along the direction of flow for both dense and sparse arrays are depicted in Figure S4. At this low heat flux condition, wedged micropillar geometries dissipates heat with comparable efficiency as cylindrical micropillars as can be seen by their average heat transfer coefficient over the entire wicking length ($\overline{h_{eff}}$). For dense array the $\overline{h_{eff}}$ values are *97.16 kW/m$^2$.K*, *88.82 kW/m$^2$.K*, and *96.96 kW/m$^2$.K* respectively for six-wedge, four-wedge and cylindrical micropillars. Similarly, for sparse array the $\overline{h_{eff}}$ values are *45.8 kW/m².K*, *43.1 kW/m².K*, and *47.5 kW/m².K* respectively for six-wedge, four-wedge and cylindrical micropillars. The above trend in $\overline{h_{eff}}$ is due to similar thermal resistance for all the micropillar geometries resulting from equivalent meniscus curvature along the wicking direction. This also results in their comparable $\Delta T_e$ values along the wicking direction as depicted in Figure S4c and Figure S4d.

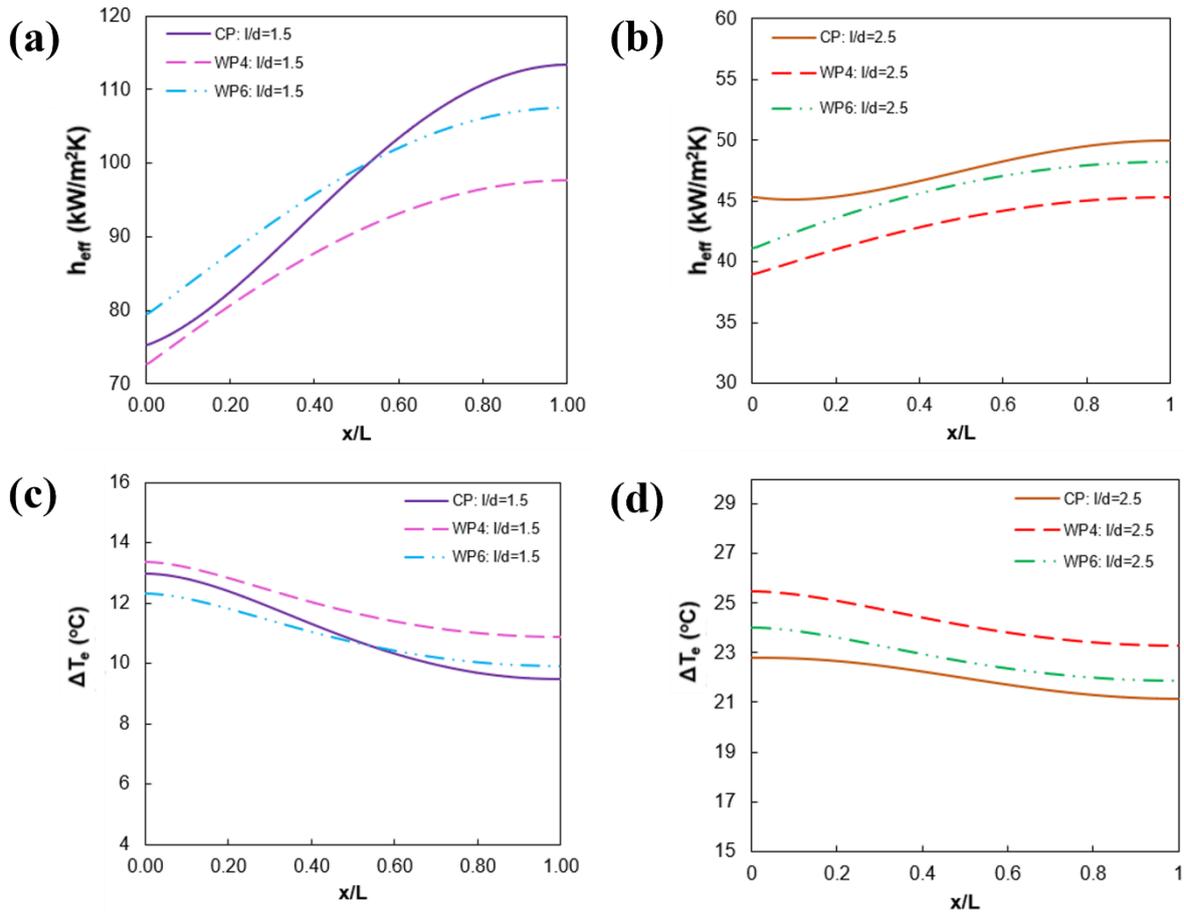

**Figure S4.** Variation of effective heat transfer coefficient for **(a)** dense and **(b)** sparse array along the wicking direction. And distribution of evaporator wall superheat for **(c)** dense and **(d)** sparse array along the wicking direction for all micropillar geometries for same heat flux of *103.5 W/cm²*, the dryout heat flux of cylindrical micropillar. Here CP, WP4, and WP6 refers to cylindrical, four-wedge, and six-wedge micropillars respectively.

**SECTION 6: Thermal resistance network**

A simplified resistance network[1] as shown in Figure S5a is considered to understand qualitatively the reasons for the improved thermal performance of biomimetic wedged micropillar geometries in comparison to cylindrical micropillar. The overall thermal resistance network can be considered to comprise of conduction resistances of micropillar ($R_P$), bulk liquid ($R_l$), and thin-film region ($R_{TF}$), and interfacial resistance across thin-film region ($R_{int,TF}$) and bulk meniscus ($R_{int,l}$).

$$R_P = \frac{L_P}{A_P k_P} \tag{S1}$$

$$R_l = \frac{L_l}{A_l k_l} \tag{S2}$$

$$R_{TF} = \frac{L_{TF}}{A_{TF} k_l} \tag{S3}$$

$$R_{int,TF} = \frac{1}{A_{int,TF} h_{lv}} \tag{S4}$$

$$R_{int,l} = \frac{1}{A_{int,l} h_{lv}} \tag{S5}$$

Here $L_P$, $L_l$, and $L_{TF}$ are the length/thickness of pillar, bulk liquid, and thin-film region respectively. $A_P$, $A_l$, and $A_{TF}$ are the cross-sectional area of pillar, bulk liquid, and thin-film region, and $k_P$ and $k_l$ are the thermal conductivities of pillar and liquid. $A_{int,TF}$ and $A_{int,l}$ are the interfacial area of thin-film region and bulk liquid meniscus. Furthermore, $h_{lv}$ is the equivalent heat transfer coefficient for evaporation calculated from Schrage's model[2]. By comparing the overall thermal resistance for both cylindrical and six-wedge micropillar unit cells at certain $\theta = 15^o$ as shown in Figure S5b and Figure S5c, we can infer that the contribution from $R_P$ increases for six-wedge micropillar unit cell due to reduced cross-sectional area for conduction heat transfer whereas all other components of thermal resistance (equation S2-S5) decreases because of (i) lower bulk liquid thickness ($L_l$) and higher cross-sectional area of bulk liquid ($A_l$) for $R_l$, (ii) higher cross-sectional ($A_{TF}$) and interfacial area of thin-film region ($A_{int,TF}$) for $R_{TF}$ and $R_{int,TF}$ respectively and (iii) higher interfacial area ($A_{int,l}$) of bulk liquid meniscus for $R_{int,l}$.

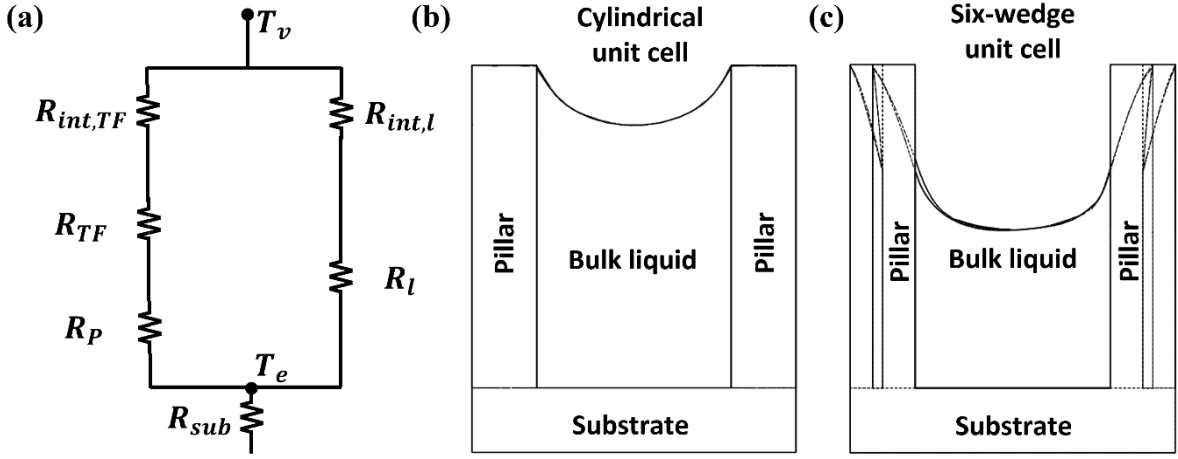

**Figure S5. (a)** Thermal resistance network representation of the micropillar unit cells to compare the qualitative thermal performance of the micropillar geometries. Unit cells of **(b)** cylindrical and **(c)** six-wedge micropillar geometries.

## SECTION 7: Laplace pressure calculation for varying contact angles along the periphery of wedged micropillars

The Laplace pressure is calculated using the force balance analysis (Equation (S6)) which takes in the wetted perimeter ($P_w$), cosine of contact angle ($\cos\theta$), surface tension ($\gamma_{lv}$), and projected meniscus area ($A_{pm}$). Conventionally, it is assumed that the contact line of the meniscus on the periphery of the micropillar maintains a constant angle throughout[3]. While this is true for a circular contact line, for the wedged micropillar geometry, as the liquid rises along the sharp corner and is pinned at the pillar top, the contact angle varies significantly along the contact perimeter as shown in Fig. S6b and this variation needs to be accounted for in the calculation of the average Laplace pressure in the unit cell. So, here we have measured the contact angle variation along the contact line by constructing planes perpendicular to the contact line and pillar wall as shown in the inset of Fig. S6a. The cosine of the contact angle is expressed as a function of the wetted perimeter (Equation (S7)) and integrated over the contact line length while calculating the Laplace pressure.

$$-P_{lap} = \frac{\gamma_{lv} \cos\theta P_w}{A_{pm}} \quad \text{(S6)}$$

$$\cos\theta\, P_w = \int_0^{P_w} f(P_w)\, dP_w \quad \text{(S7)}$$

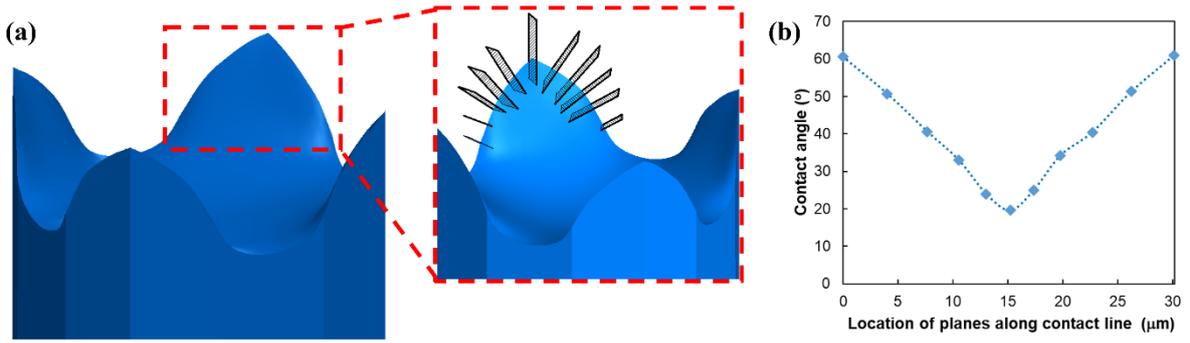

**Figure S6. (a)** Liquid meniscus formed within a unit cell of four-wedge micropillar with $\theta=15^\circ$. The construction of various planes is highlighted in the inset; **(b)** Contact angle variation along the contact line of the liquid meniscus formed within a unit cell of a wick with micropillars with four-wedges.

## SECTION 8: Grid independence check

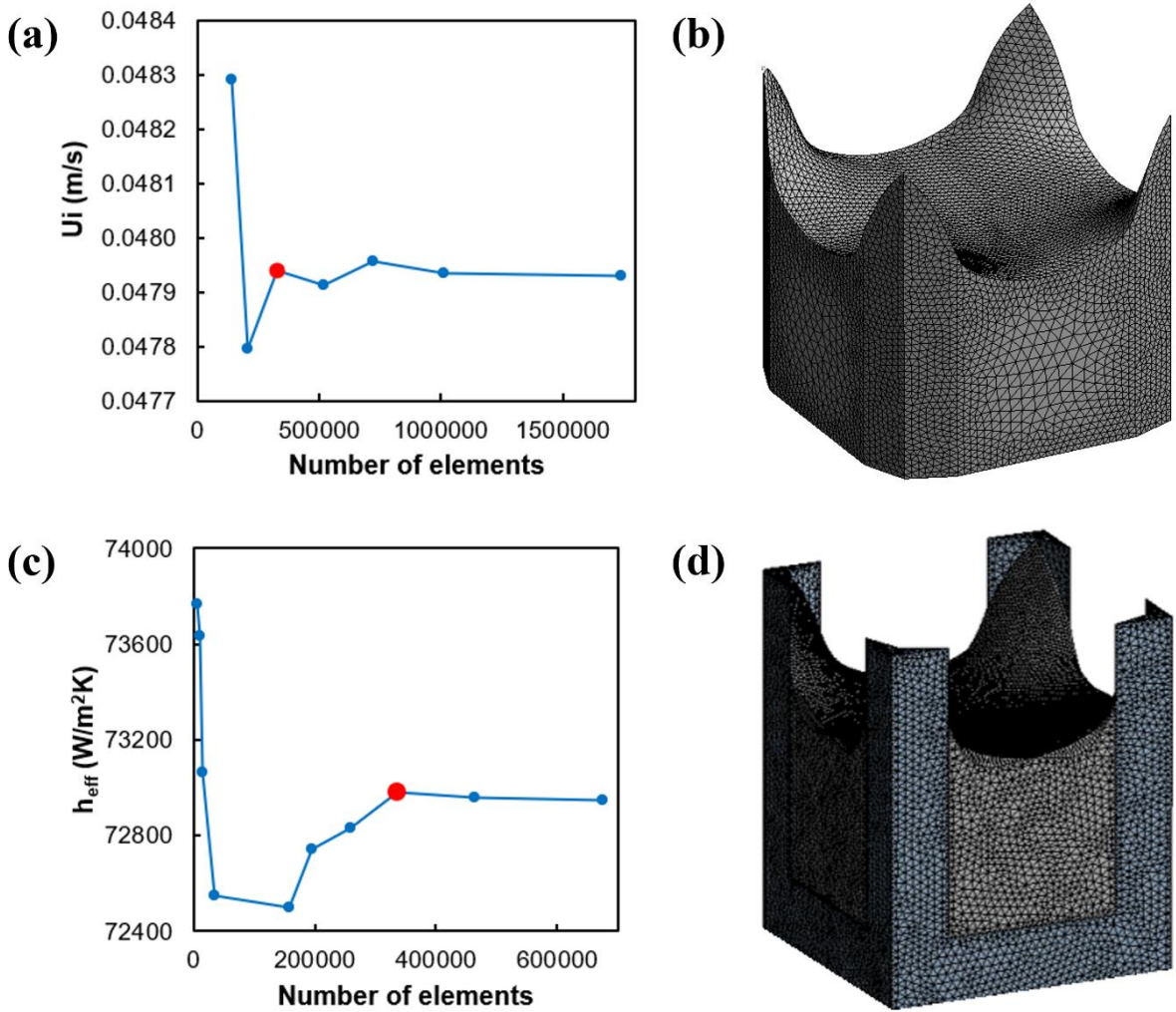

**Figure S7.** Grid independence check demonstrated with **(a)** midplane averaged velocity and **(b)** selected mesh domain for the fluid flow simulation; Grid independence check demonstrated with **(c)** effective heat transfer coefficient and **(d)** selected mesh domain for the steady state thermal simulation; for the case of four-wedge micropillar unit cell with *d=20 μm, l=50 μm, h=50 μm, and θ=15°*

The effect of mesh refinement on the fluid flow and steady-state thermal simulation is detailed here for a four-wedge micropillar unit cell with *d=20 μm, l=50 μm, h=50 μm, and θ=15°*. Figure S7a depicts the midplane averaged velocity of the unit cell as a function of the number of elements for an applied pressure gradient of *0.3 Pa/μm*. The selected mesh (data point marked red in Fig. S7a) is optimal at *326106* elements for an average element size of *8.5E-7 m* as the result changes within the limit of only *0.1%* compared to the next level of mesh refinement. The meshed domain used for fluid flow simulation is as shown in Fig. S7b.

Similarly, Fig. S7d shows the meshed domain used for steady state thermal simulation and Fig. S7c shows effective heat transfer coefficient calculated for a unit cell as a function of the number of elements for an applied heat flux of *500 W/cm²*. Here also the optimal mesh (335204 elements at 7E-7 m of average element size) is selected (data point marked red in Figure S7c) such that further refinement does not change the result. Similarly, the grid independence check is performed for all the unit cells corresponding to other contact angles.

**SECTION 9: Unit cell and device level heat transfer model**

A steady-state thermal simulation is conducted for each unit cell to determine the effective heat transfer coefficient ($h_{eff}$) due to evaporation based on the input heat flux ($q''$) and the evaporator wall superheat ($\Delta T_e = T_e - T_{sat}$) as provided by equation (7) in the main text. Convection on the liquid-vapour interfacial meniscus (yellow), symmetry on lateral sides (dark blue), adiabatic on the exposed pillar (red), and heat flux at the substrate's bottom face are among the boundary conditions (Figure S8a) applied in the unit cell steady state heat transfer model. As detailed in the main text, $h_{eff} = f(P_{lap})$ is coupled with device level model for dryout heat flux to determine the effective heat transfer coefficient distribution along the micropillar array for any given geometry. This variation of effective heat transfer coefficient along the wicking direction ($h_{eff} = f(x)$) is applied along with other boundary conditions as shown in Figure S8b on the entire array of length *L* to determine the distribution of evaporator base temperature ($T_e$).

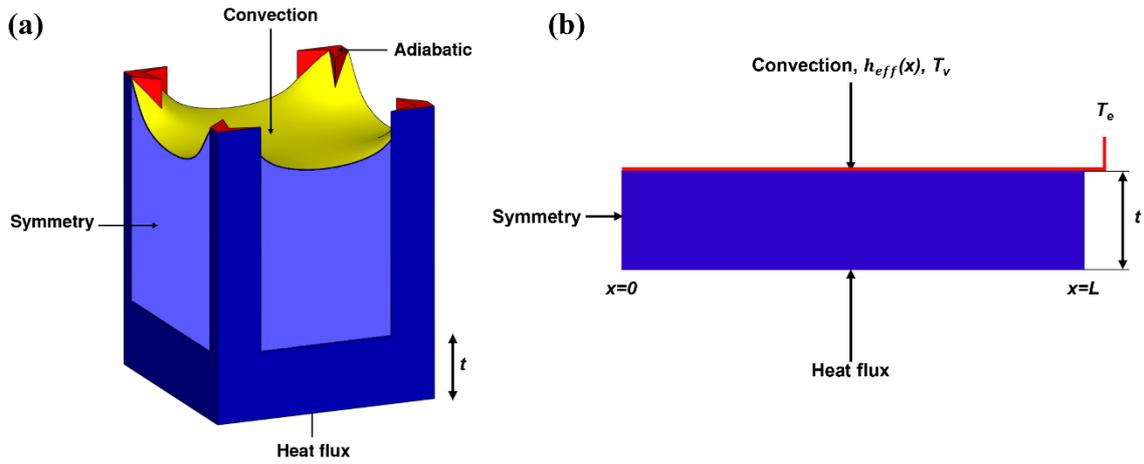

**Figure S8. (a)** Boundary conditions applied in the unit cell steady state heat transfer model include convection on the liquid-vapour interfacial meniscus (yellow), symmetry on lateral faces (dark blue), adiabatic on exposed pillar surfaces(red), conjugate heat transfer condition on pillar surfaces in contact with liquid and heat flux at bottom face of the substrate; **(b)** Boundary conditions applied in the device level steady state heat transfer model include convection on the evaporator base, symmetry on lateral faces, and heat flux at bottom face of the substrate.

**SECTION 10: Validation of the numerical model**

The present numerical model is validated by considering a cylindrical micropillar geometry of *d=10 μm, l=30 μm, h=25 μm, t=100 μm, and L=5 mm*. The variation of Laplace pressure ($P_{lap}$) along wicking direction (*x/L*) is corroborated against the data of Zhu et al.[3] as shown in Figure S9a and found to be in good agreement, justifying the Surface Evolver-based modeling approach's excellent prediction of interface shape. The dryout heat flux predicted using the current model was found to be *81.76 W/cm$^2$* and that from the Zhu et al.[3] and Vaarstra et al.[4] models were *76.1 W/cm$^2$* and *82 W/cm$^2$* respectively. This again substantiates the accuracy of the current modeling approach. We have also validated the result of the effective heat transfer coefficient ($h_{eff}$) distribution along the wicking direction for an applied heat flux corresponding to that of the dryout heat flux as shown in Figure S9b. And, found that the current model gave an average $h_{eff}$ of *76.88 kW/m$^2$.K* whereas the Vaarstra et al.[4] model's was *75.5 kW/m$^2$.K* reiterating the excellent accuracy of the Surface Evolver based numerical modeling methodology.

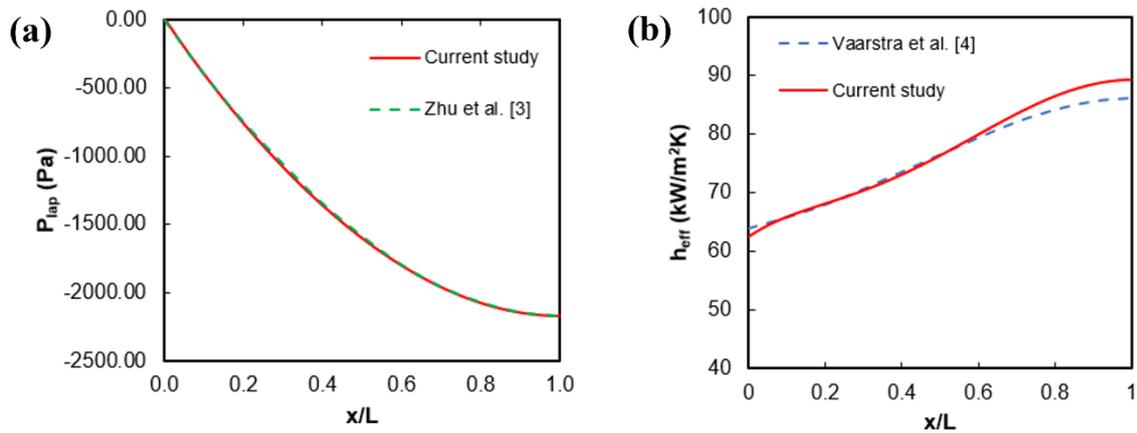

**Figure S9.** Validation of the **(a)** Laplace pressure, **(b)** effective heat transfer coefficient along the length of micropillar array with that of the literature.